A Chapter in *Polymer Rheology*

# Rheological Properties of Thermoplastic Polymers with Dissolved Gases for Foaming Applications

Daniel Raps*, Department of Polymer Engineering, University of Bayreuth, Bayreuth, Germany, daniel.raps@uni-bayreuth.de

Hossein Goodarzi Hosseinabadi, Polymeric Materials Research Group, Department of Materials Science and Engineering, Sharif University of Technology, Tehran, Iran, hgoodarzy@gmail.com

Lutz Heymann, Department of Applied Mechanics and Fluid Dynamics, University of Bayreuth, Bayreuth, Germany, lutz.heymann@uni-bayreuth.de

Thomas Köppl, Department of Polymer Engineering, University of Bayreuth, Bayreuth, Germany, thomas.koeppl@uni-bayreuth.de

Volker Altstädt, Department of Polymer Engineering, University of Bayreuth, Bayreuth, Germany, altstaedt@uni-bayreuth.de

## Abstract

Understanding the concept of rheology in polymer melts with dissolved gases like $CO_2$ is crucial in development of high quality polymer foams which are commonly manufactured by foam extrusion processes. The crystallization and rheological properties of the melt are significantly affected by the dissolved gas, which acts as a blowing agent and at the same time as a plasticizer. Moreover, the rheological properties of gas-loaded polymer melts influence on the efficient design of extrusion dies toward production of popular homogenous lightweight foams. This chapter will discuss common methods to measure these rheological properties with the example of a long-chain branched Polypropylene as a case study. To take into account the role of crystallization behavior under gas-loading into the process design, we describe a method to measure both rheology and crystallization in one experimental setup.

## 1. Introduction

Controlling the rheology has been a technological bottleneck in manufacturing polymeric foams, which are widely produced by extrusion processes. Among materials of utmost interest in the foam industry, thermoplastic polymers such as Polypropylene (PP) are outstanding choices due to their favourable mechanical properties and chemical resistance. However, their complex microstructure and semi-crystalline nature result in unfavourable temperature dependence of their low melt elasticity and viscosity, which makes it difficult to obtain homogeneous lightweight foams. A good control over the rheology is therefore essential to adjust the morphology and thus the elastic modulus, strength [1], impact behaviour [2], and thermal conductivity [3] of foamed products.

During foam extrusion, temperature, pressure, and blowing agent affect the flow behaviour of the melt and ultimately the foam properties. In addition, the melt behaves differently under shear and elongational flow. The flow patterns in the extruder are dominated by shear deformation. However, the elongational properties become relevant after the melt leaves the die, since in foaming stage the melt is subjected to elongational flow during bubble growth, namely equi-biaxial extension. This chapter gives a brief overview of the theoretical aspects and concepts of foam processing, followed by an overview of methods to evaluate the rheology of gas-loaded melts. Ultimately, exemplary results obtained from two distinctly different methods are shown for a foaming grade PP (i.e. high melt strength (HMS) PP).

## 2. Theoretical background

Flow-curves of polymer melts under various conditions exhibit self-similarity if their relaxation mechanisms have an identical dependency on environmental parameters. This phenomenon leads to the concept of reduced variables as described by Ferry [4] and based on a thermodynamic model [5]. In this situation, the viscosity curves at different conditions can be shifted to a master-curve by applying horizontal and vertical shift factors. The contribution of horizontal shift factor $a$ into the shear viscosity of the melt is applied in the following way:

$$\eta(\dot{\gamma}) = a \cdot \eta\left(\frac{\dot{\gamma}}{a}, T_{ref}, p_{ref}, c_{ref}\right) \quad (1)$$

Here, the index *ref* denotes reference states of temperature $T$, pressure $p$ and gas concentration $c$. $\dot{\gamma}$ stands for the shear rate, $\eta$ for the shear viscosity. In this formulation, horizontal shift factors for the effect of temperature $a_T(T)$, pressure $a_p(p)$ and gas-concentration $a_c(c)$ can be combined to the general horizontal shift factor in the following way:

$$a = a_T(T) a_p(p) a_c(c) \quad (2)$$

The temperature dependency of the shift factor $a_T$ for temperatures above 100 K over the glass transition temperature can be described by an Arrhenius dependency (3), which is based on the idea of an energetic barrier for slip between polymer chains [6]. For a reference temperature $T_{ref}$ and a process temperature $T$, the Arrhenius-equation for $a_T$ then yields:

$$\ln(a_T) = \frac{E_a}{R}\left(\frac{1}{T} - \frac{1}{T_{ref}}\right) \qquad (3)$$

with an activation energy for flow $E_a$ and the universal gas constant $R$.

In contrast to the effect of temperature on shear viscosity, an increase of pressure leads to an increase of viscosity. For processes that are running at high pressure, like injection moulding and in a lesser degree extrusion, this is an important factor. Therefore, it is of utmost importance to measure and quantify the effect of pressure on viscosity variations. The increase of viscosity is caused by a decrease of free volume with increasing pressure and hence a reduced segmental mobility of polymeric chains. This is expressed by the Barus-equation [7]:

$$\ln(a_p) = b(p - p_{Ref}) \qquad (4)$$

Where $\beta$ denotes the isothermal compressibility of the melt.

Like temperature and pressure, dissolved gas concentration has a distinct effect on the polymer processing. According to Park and Dealy [8], the decrease of viscosity is dependent on the number of gas molecules dissolved in the polymer network. Once again, the free volume theory can be used to describe the decrease in viscosity [9]. The dissolved gas not only acts as a blowing agent in the foaming process, but also contributes like a plasticizer in the system. Besides changing the rheological behaviour, dissolved gas also leads to a decrease of the crystallisation temperature $T_C$ [10] and glass transition temperature $T_G$ [11]. Furthermore, the crystallisation kinetics are affected as well [12,13].

On the other side, thermo-rheologically complex materials often show an influence of temperature on the density and, hence, on the rheological properties of the material. Therefore, assuming for un-entangled polymers a proportionality between stress and the product of temperature $T$ and density $\Gamma$ [5] (or for entangled polymers a proportionality between the relaxation modulus and the product $k(T) \cdot \rho \cdot T$), a vertical shift factor can be introduced by the relation:

$$b_T = \frac{T_{ref} \cdot \rho_{ref}}{T \cdot \rho} \qquad (5)$$

where $\rho$ denotes the density under experimental conditions and $\rho_{ref}$ demonstrates the density under reference conditions. Since the dimensionless $b_T$ only varies in the order of 1 by a few percent, in practical application it is often set as $b_T = 1$.

In contrast to the shear deformation, the elongational properties of the melt govern the cell growth mechanism and thus influence the rupture of the forming cell walls. Reliable data on the rheology of the gas-loaded melt are therefore required to accurately control the foaming process. Such data are also necessary as input for simulations, e.g. for optimization of the die geometry used for foaming. To reduce development costs of foam injection moulded parts, simulations can help to avoid errors in early development stages [14]. The viscoelastic behaviour of the melt in elongational deformation is crucial for the formation of the foam morphology, as mostly low average cell size, a narrow cell size distribution, and low density are required. In particular, the effect of strain hardening at large elongational strains is desired to support cell stabilisation. Strain hardening means the rise of the elongational viscosity $\eta_E(t)$ above the zero-rate elongational viscosity $\lim_{\dot{\varepsilon}\to 0}\eta_E(\dot{\varepsilon})$. It helps to prevent cell coalescence and to widen the processing window [15], i.e. for the foaming of PP [16]. In terms of chain topology, strain hardening is caused by long chain branching (LCB) [17,18]. However, LCB-PP has the disadvantage of a lower solubility of carbon dioxide ($CO_2$) compared to linear PP [19]. This arises due to the fact that LCB-PP has a lower specific volume compared to linear PP as well as a more pronounced resistance against swelling, which is caused by dissolution of $CO_2$ [20].

Compared to linear polymers the rheological characterisation of LCB polymers is more challenging, since they possess characteristics, which complicate the measurements and analyses. For example, LCB polymers may not obey the Cox-Merz empirical rule which states that the dependence of the *steady* shear viscosity on the shear rate can be estimated from the *dynamic* viscosity of the melt [21–23]. Note that it is too difficult to obtain the zero-shear viscosity of heavily long-chain branched polymers from oscillatory data. Their very long relaxation time would require very low frequencies, which cannot be accessed experimentally without thermal degradation of the material. Hence, creep experiments must be performed to obtain this material's property [24].

A further challenge is thermo-rheological complexity of LCB polymer melts. It is known that the time-temperature-superposition (TTS) of many polymer melts can be described by only one pair of shift factors per temperature, leading to a so-called "master-curve". This behaviour is then called thermo-rheologically simple. If it is not possible to obtain a satisfactory master-curve, as a systematic deviation with temperature is observed (e.g. the shift factors are dependent on stress, modulus or relaxation strength), the polymer system is then called thermo-rheologically complex. In thermo-rheologically complex polymers, two or more relaxation mechanisms do not have the same temperature dependency [25], which leads to time or frequency dependent temperature shift factors. LCB often causes thermo-rheological complexity, which is shown by several authors [26–28].

## 3. Experimental evaluation of the rheological properties of gas-loaded polymer melts

For the challenging task of measuring gas-loaded melts special experimental setups are devised. Some important and widely used setups are reviewed here.

### 3.1 Pressure Cell

For measuring shear properties there exist several designs of devices. A good method is a air-tight pressure cell. Beside polymers, the pressure cell was used to study the flow properties of a foamed oil [29]. Few researchers also applied it for polymer melts. Flichy et al. [30] investigated the viscosity of silica-particle filled Polypropylene glycol (PPG) with such a device. Using a gas with high solubility and diffusion rate, the sample can be saturated during the measurement. Later, Wingert et al. [31] studied the effect of $CO_2$ on the viscosity of Polystyrene (PS) using a Couette-geometry.

To bring down the very long diffusion times (axial diffusion, circa 1 year saturation time) to a bearable time scale, a porous cup was used.

Hangde and Altstädt [11] also investigated the influence of pressure and $CO_2$ on PS melts. In contrast to the previous publications, they used a plate-plate geometry. To minimize diffusion time, they compared samples that were preloaded with $CO_2$ in an autoclave with unloaded samples. The preloaded samples reached equilibrium much earlier. They also compared the device to a standard rheometer and found that the results overlap well in both sets of devices. The pressure cell is built of a gas-tight measurement cell of titanium as is shown schematically in **Figure 1**. It is designed to keep a maximum temperature of 200 °C and a maximum pressure of 400 bar. In this cell, the upper moving plate is hold by ball-bearings. The torque is transferred magnetically between a magnetic coupling fixed to the rheometer measuring head and magnets inside the cell. According to Handge et al. [11] the accuracy of the cell is high for sufficiently viscous polymers when a plate-plate geometry is used.

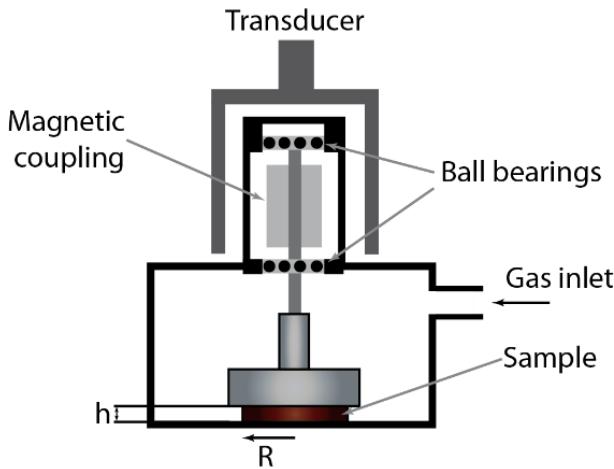

**Figure 1: Pressure cell for measurements under high hydrostatic pressure using a plate-plate geometry.**

The procedure and governing equations for this device are shown below as its results will be shown in the 4[th] part of this chapter as exemplary results.

First, the sample is introduced into the hot cell/chamber. After an equilibration period of four minutes that equilibrates the sample in a melt phase, the sample is subjected to a constant shear-rate (no oscillation) with uniform rotation of the upper plate. After 120 s of shear deformation, a certain pressure of $CO_2$ is applied into the chamber. The gas can be delivered from a syringe pump, capable of delivering a maximum pressure of 400 bars. After 300 s, a rest period at a shear-stress of $\tau = 0\ Pa$ is applied to give the polymer some time for relaxation. An iteration of start-up tests, which is a sudden start of a defined shear-rate, and the rest period is conducted till the equilibrium of viscosity is achieved. This is done for each individual $CO_2$ pressure and temperature of interest. This step allows the study of diffusional properties of $CO_2$ in a long chain branched (LCB) polymer.

After the diffusion step, the flow curve of the melt is measured which then allows the investigation of gas loading on the flow curve of the melt and finally the calculation of the relevant pressure and gas-concentration shift factors. Shift factors are obtained from shifting the viscosity-curves over each other, according to equation (1). Accordingly, the viscosity can be calculated from torque M. A sinusoidal torque offset s function of deflection angle $M(\phi(t))$ due to the ball-bearing must be considered. Also, the radius of the sample depends on the gas-concentration and thereby the saturation pressure. The sample radius, R, can be calculated using the gas-free density at a certain pressure and temperature, $\rho_{melt}(T,p)$, as well as the swelling ratio, $Sw$, from material data according to Li and Park [20] equation:

$$R = \sqrt{\frac{m\ Sw}{\rho_{melt}(T,p)\ h\ \pi}} \quad (6)$$

with sample mass, $m$, and gap height, $h$. The gap height is conventionally considered 0.8 mm. Each sample is weighted on a high precision scale before running the experiment. Using this equation for a plate-plate geometry, the governing equations for shear-rate and stress yield:

$$\dot{\gamma}_R = \omega \sqrt{\frac{m\ Sw}{\rho_{melt}(T,p)\ h^3\ \pi}} \quad (7)$$

and

$$\tau_{app} = \frac{2(M - M(\phi(t)))}{\left(\frac{m\ Sw}{\rho_{melt}(T,p)\ h\ \pi}\right)^{3/2}} \quad (8)$$

with angular speed, ω, sample mass, $m$, and swelling ratio, $Sw$, due to $CO_2$-diffusion.

Finally, the third step to investigate the crystallisation behaviour under high hydrostatic pressure and gas-loading is conducted. To this aim, the pressure cell is cooled down from the measurement temperature of 180°C at a constant rotational speed of 0.19 rpm. When crystallization takes place, the viscosity of the sample increases dramatically till a maximum is reached, as can be seen in **Figure 2**. At this maximum the solidified sample is loosening from the rheometer surface and then slipping, so only friction is responsible for torque beyond that point. In this situation, the onset-temperature of crystallization is chosen to be the temperature, where torque deviates more than 10 % of its average baseline before the torque-maximum. It is evaluated by firstly creating a baseline before crystallization by fitting the data from the start to the onset of the torque-peak, and then this line is raised by 10 % offset. Finally, the temperature at the crossover-point between this 10 %-line and the data is considered as the temperature of the onset of crystallization. This procedure is graphically shown in **Figure 2**.

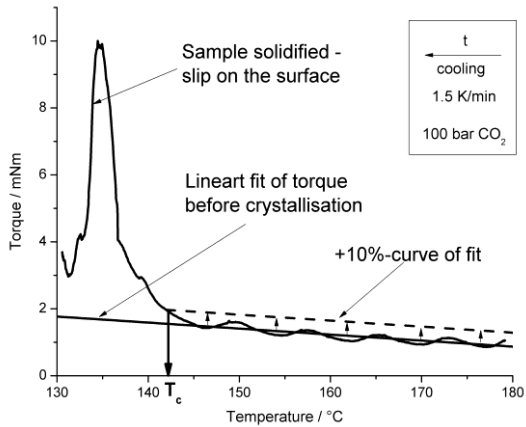

**Figure 2:** The evaluation procedure for crystallization: the baseline before crystallization is raised by 10 % offset, the temperature at the crossover point between this 10 %-line and the data is defined as the onset of crystallization temperature, $T_C$.

### 3. 2. In-Line rheometry

In-line rheometry for foam extrusion lines is a commonly used method to obtain viscosity data under gas-loading [32–37]. Shear rate rates between 10 and $10^3$ are usually accessed depending on the setup. This setup consists of an extruder or a tandem extrusion line, a system for $CO_2$-delivery, and a slit or capillary die. A back-pressure assembly must be mounted on the die, too, to assure a pressure above the critical solubility pressure (under the so-called supper critical condition). The advantage of an in-line rheometery is the separability of the pressure- and gas dissolution-effect by running experiments at different combinations of gas-loadings and pressures. Optionally a melt pump can be used in front of the die to deliver a constant volume flow. Furthermore, utilizing a heat exchanger over the in-line rheometer can give the benefit of better melt-temperature stability.

Measuring the shear properties of a gas-loaded melt is established in literature as shown previously by numerous references, although the absolute number of publications is comparatively small to other fields such as the influence of processing parameters on foam structure. The situation is worse for elongational measurements, as only few works could so far investigate the elongational properties of gas-loaded polymer melts. One possible method applies the Cogswell approach [38] by using a sealed capillary rheometer or an in-line rheometer [34]. Another method uses hyperbolically converging dies on an in-line rheometer. A first attempt with gas loaded melts on this equipment was reported by Wang et al. [39,40], where they calculated the elongational viscosity from the pressure drop along the hyperbolic die. The obtained data of unloaded melts were in good agreement with the data measured using commercially available elongational rheometers. The mentioned approaches apply a mathematical procedure to decompose the complex flow into its shear and elongational components. However, the success of such a procedure cannot be assured, as this decomposition might not yield correct results. Thus the results should be evaluated carefully.

For the study in part four in this chapter, a rheological die is applied in order to measure the shear viscosity in-line during the foam extrusion process. The setup is schematically shown in **Figure 3**-a), the design of the die is shown in **Figure 3**-b), and the shear properties are measured with a slit insert. Optionally, the elongational properties can be explored using a hyperbolically converging die insert. This option is not shown here and can be found in the original contribution [41]. Both shear and elongational viscosities are calculated from the pressure drop along the respective die, which is measured with calibrated pressure transducer.

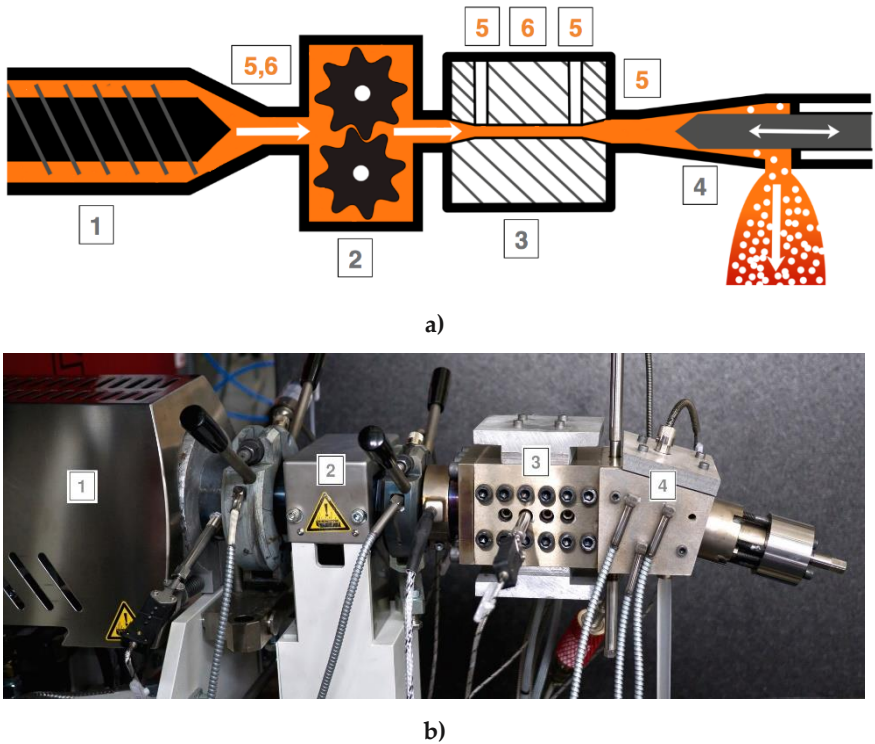

**Figure 3: a) Sketch and b) picture of the in-line rheometer: 1: single screw extruder in tandem line, 2: melt pump, 3: slit die (alternatively hyperbolic die), 4: back pressure assembly, 5: pressure sensors, 6: temperature sensors.**

Two heating plates on top and bottom of the die control the temperature inside the measurement section. Their heating power is set by a control loop. The necessary temperature information is obtained from a temperature sensor in the die. The temperature of the backpressure assembly is similarly controlled and instead of heating plates, heating rods are used in the setup.

### 3. 3. Further methods

To obtain data for higher shear-rates, relevant for polymer processing, capillary rheometry is an appropriate method. Gerhard et al. [42] for the first time used a sealed capillary rheometer to measure $CO_2$ loaded Polydimethylsiloxane. The sample was saturated in an autoclave and transferred to the rheometer by means of a high-pressure screw pump.

Later, Kwag et al. [43] used a sealed capillary rheometer. They loaded polystyrene granules at high pressure in an autoclave with different gases until the desired gas-concentration was reached (polymer not molten!). This was checked with an electronic balance. After saturation the pallets loaded with $CO_2$ were quenched in liquid nitrogen to minimize the gas loss. To keep pressure up during melting the piston applied a constant pressure on the pallets while an outlet-valve was closed.

A method to achieve even higher shear-rates up to $10^5$ $s^{-1}$ is to operate a special die on an injection molding (IM) machine, which is introduced by Qin et al. [44] The IM-machine was operated with a polymer/chemical blowing agent (BA) mixture. The chemical BA degrades during the filling stage of the IM operation and releases gas. Due to high pressure the gas dissolves in the polymer. The die consists of an adapter to mount it onto the IM-machine, a standard capillary channel and two pressure and temperature transducers respectively. Two capillaries are used to correct the entrance pressure loss. The results obtained by this method are reported to be in agreement with off-line rheometry methods.

## 4. An exemplary study on the gas-loaded rheology of long-chain branched Polypropylene

The following section presents a selection of results of previous studies on a foaming-grade polypropylene [41,45]. The rheological properties of gas-loaded high melt strength Polypropylene (LCB-PP, Daploy WB140HMS from Borealis, Austria) with a MFI of 2.1 g/10min (230 °C, 2.16 kg) [16] is studied. $CO_2$ is used as the blowing agent.

### 4.1 In-line rheometry

The shear properties of $CO_2$-loaded LCB-PP are measured in the in-line rheometer shown in **Figure 3** at a pressure of 150 bar. **Figure 4** shows the effect of different gas-concentrations on shear viscosity at 180 °C. At 6 % $CO_2$ concentration, the shear viscosity is reduced by 70 % compared to the unloaded sample at the same shear rate. The $CO_2$-concentration of 6 % in the melt at 180 °C corresponds to the same viscosity as a temperature increase of 40 °C of the unloaded melt.

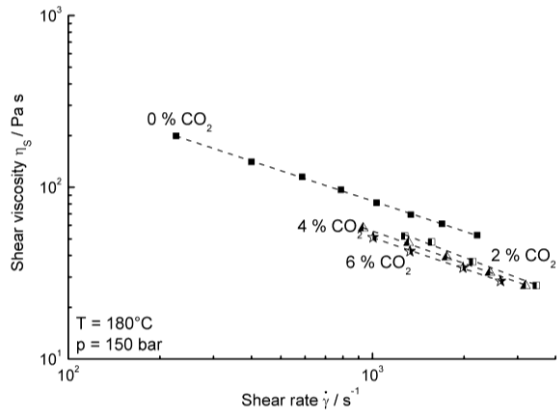

**Figure 4: Effect of different $CO_2$-loadings on the shear viscosity function at 180°C and 150 bar (in-line slit-die experiment).**

It is observed, that dissolved gas exhibits a pronounced effect on the viscosity function due to plasticisation. The large difference between the gas-free melt and a melt with 2 % $CO_2$-loading (**Figure 4**) suggests, that small amounts of gas measurably reduce the viscosity. This effect was also observed in other studies [32,35]. The dissolved gas molecules can reduce the friction between the polymer chains and increase the free volume thus reducing viscosity.

The inter-dependency of temperature and the amount of dissolved gas is also investigated. To show the influence of gas loading, the shift factors for dissolved gas $a_C$ were calculated for various temperatures. The shift factor attains smaller values for a distinct gas-effect and vice versa. The concentration shift factors $a_C$ for various temperatures and $CO_2$-concentrations are shown in **Figure 5**. At high temperature, the gas-effect is less pronounced than at lower temperatures, which can be seen from the increase of the shift factor when increasing temperature.

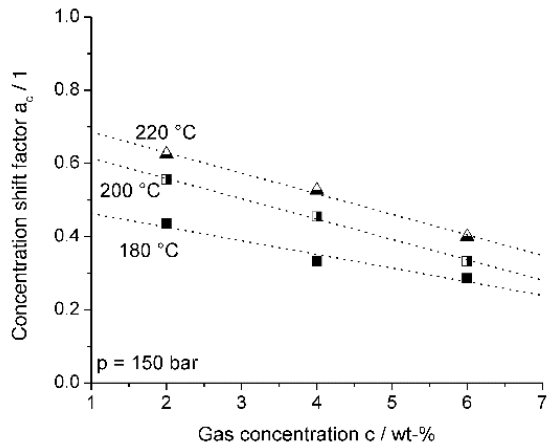

**Figure 5: The effect of $CO_2$-loading on the concentration shift factor $a_C$ and its interplay with melt-temperature.**

At a constant temperature, the concentration shift factor $a_c$ decreases with increasing gas concentration, which corresponds to an increased plasticising effect ($a_c < 1$). This was also observed by other researchers [8,46,47], who related this observation to the free volume theory.

The gas effect is less pronounced at elevated temperatures than at lower temperatures. The same effect was reported by Royer [46]. He observed a temperature dependent concentration shift factor ($a_c$) for polystyrene (PS) and related this finding to the free volume theory. Also, Kwag et al. [48] reported a similar effect for PS. However, the free volume cannot be responsible for this behaviour, since it does not dominate the viscosity far above the glass transition temperature, $T_G$. It is hypothesised that the potential for plasticisation, due to dissolved gas, decreases at higher temperatures. Hence, the effect of plasticisation is expected to be smaller at higher temperatures.

In summary, it was seen that the dissolved gas has a pronounced influence on the viscosity as measured in an in-line rheometer, as well as its dependency on the deformation rate and temperature. The viscosity was reduced by 70 % at a $CO_2$ content of 6 wt-% for a constant shear rate, which equals with an increase in temperature of 40 °C for the unloaded melt (reference temperature 180 °C). This phenomenon emphasizes the importance of gas loading on the rheological behaviour of the melt. For example, the depletion of $CO_2$ in the melt during bubble growth leads to an increase of viscosity, which supports the stabilisation of the cellular morphology in a polymeric foam product.

Although these results are very promising, the in-line technique also has some serious drawbacks. First of all, significant quantities of material are required (depending on extrusion line > 100 kg) to cover a broad range of shear rates and tests for reproducibility examination. Thus, the in-line method is rather limited for the use of small-scale material development. Furthermore, care has to be taken when interpreting the results on pressure dependency. If the polymer has very long relaxation times, as in the case of the studied HMS-PP, then very low pressure-sensitivities are measured if the pressure in the extruder is significantly lower than in the slit section. Also dissipative heating might be an issue, if high shear rates corresponding to high melt pump *rpm* (in this case > 50 rpm) are tested. Thus, a temperature sensor is required to check, whether the temperature increased significantly above the desired value. Also, the amount of invest is rather high as a foam extrusion line with a gas-dosing system is a requirement.

### 4.2 Pressure cell

When diffusion of $CO_2$ into the sample is finished, the flow curves can be acquired from consecutive dynamic/transient experiments. The resulting steady-state viscosities are plotted in **Figure 6**. The viscosity reduction by dissolved $CO_2$ is eminent. A viscosity reduction of *one order of magnitude* at a constant shear-rate in the non-linear region at a saturation pressure of 400 bar can be seen compared to non-dissolved gas samples. The Henry equation is then used to calculate the solubility of the gas inside the melt. It reads:

$$S = k_H p \qquad (9)$$

which relates the gas solubility, *S*, to the gas pressure, *p*, *using* the Henry constant, $k_H$. The Henry coefficient of LCB-PP is $k_H$ = 8.5 wt%/100 bar at 180 °C and $k_H$ = 7.5 wt-% / 100 bar at 200 °C [19]. Thus, saturation pressure of 400 bar is equivalent to a gas-concentration of $c_{(CO2)}$ = 34 wt-% according to Henry's law. Note that at gas pressure of 400 bar, the measurement of very low shear-rates are experimentally not accessible due to too small torque values.

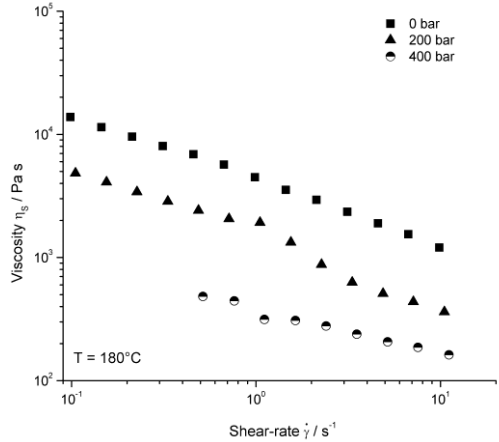

**Figure 6: Flow curves at different saturation pressures of CO2 at 180 °C.**

By shifting the viscosity curves over each other, the combined pressure and gas-concentration shift factor can be deducted as is plotted in **Figure 7**, where the viscosity decrease in a linear region can be seen. Interestingly, zero shear-viscosity is deceased by a factor of 300 ($1/a_{P,C}$) at saturation pressure of 400 bar (equivalent to $c_{(CO2)}$ = 34 wt-%).

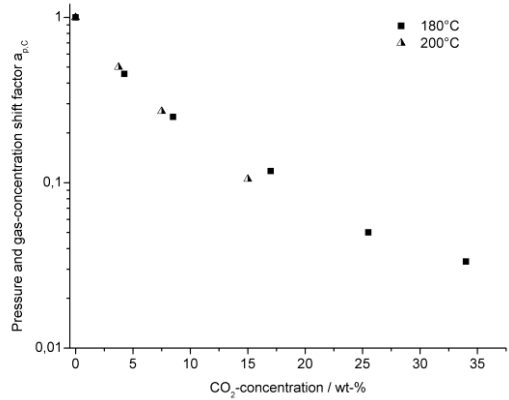

**Figure 7: Pressure and gas-concentration shift factor dependence on the amount of dissolved CO$_2$.**

The non-linear behavior (the change of slope with increasing gas-concentration) is caused by the non-linearity of the pressure effect on viscosity. By using point-wise pressure shift factors, the gas-concentration shift factor can be separated from the pressure effect, as depicted in **Figure 8**. The gas-concentration shift factor shows an exponential decrease with gas-concentration.

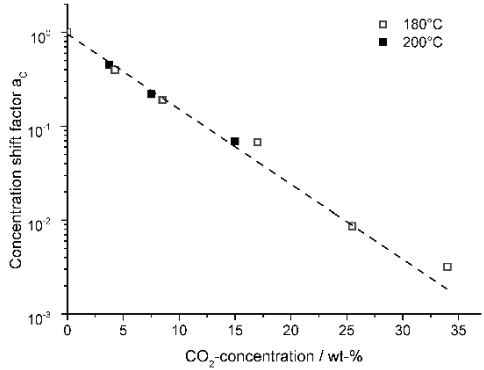

**Figure 8: Gas-concentration shift factor dependent on the amount of dissolved $CO_2$.**

It is clear that the viscosity is reduced significantly by introduction of $CO_2$ to the system. In other words, the effect of pressure is more significant than being balanced by the effect of dissolved gas concentration.

### 4. 3. Influence of dissolved gas on the crystallisation temperature

Torque-temperature curves at different saturation pressures are presented in **Figure 9**. From this plots it is clearly observable that crystallization is delayed, when the amount of dissolved $CO_2$ increases in the polymer.

The obtained crystallization temperatures, $T_{Cs}$, are plotted as a function of saturation-pressure in **Figure 10**. It can be observed that $T_C$ effectively decreases with increasing saturation pressures. This could be probably due to the increased free volume and thus the chains mobility when gas is present [49]. At high pressures the slope of the curve is levelling off due to an increased pressure effect at high pressures. The equivalent effect was also seen for viscosity dependency on pressure, where pressure effect increased above 200 bar as described previously.

Crystallisation temperature in dependency of $CO_2$ pressure was fitted by a 2$^{nd}$ order polynomial which reads: $T_C = T_0 + B_1 p + B_2 p^2$. This non-linear fit is used because compressibility is decreasing with increasing the pressure [50]. The parameters determined by this fit for LCB-PP are $T_0 = 428.31 \pm 1.5$ K, $B_1 = -0.115 \pm 0.019$ K/bar and $B_2 = 8.682 \cdot 10^{-5} \pm 4.714 \cdot 10^{-5}$ K/bar$^2$. The fit had an $R^2$-value of $R^2=0.981$ indicating a satisfactory fit quality. At a saturation pressure of $p = 0$ bar, the pressure sensitivity yields $B_1(0 \text{ bar}) = -0.115$ K/bar, while at $p = 300$ bar pressure the sensitivity is only $B_1(300 \text{ bar}) = -0.0889$ K/bar.

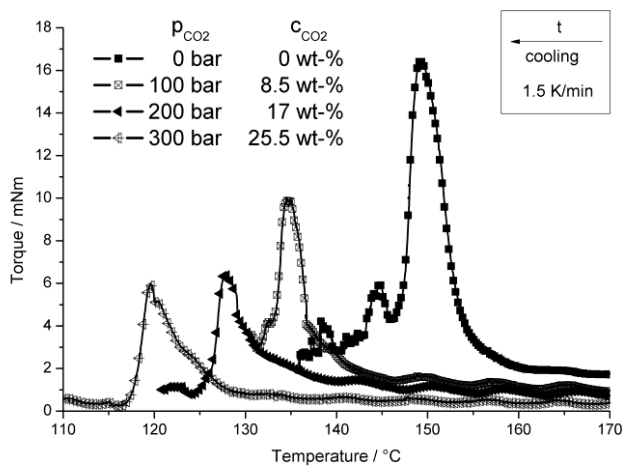

**Figure 9: Crystallisation process at different saturation pressures at a cooling-rate was 1.5 K/min.**

In literature, values of the same order of magnitude were reported for PP: Takada et al. [51] found a dependency of 0.275 K/bar for isotactic PP ($M_W$ = 85,000 g/mol, PDI 3.3) using a high-pressure differential scanning calorimeter (HPDSC). The lower value can be explained by the different molecular structure of the material, which has no long chain branching, and a different measurement method. Naguib et al. [52] investigated the effect of pressure and dissolved gas on crystallization for linear and LCB-PP using HPDSC. For low pressures below 60 bar he found a drop in $T_C$ of 0.3 K/bar. In his study a non-linearity at low pressures was found.

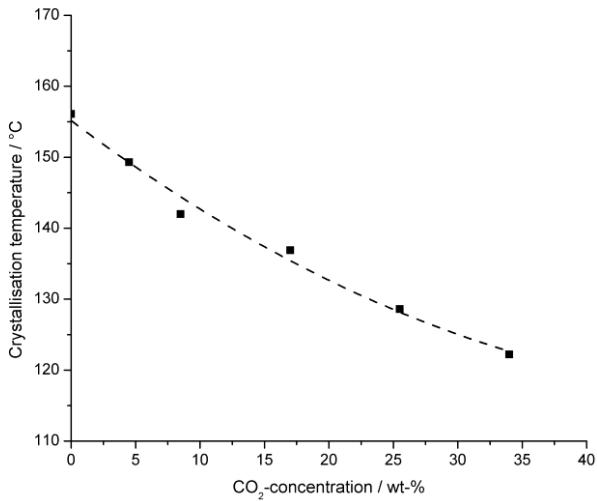

**Figure 10: Crystallisation temperatures as function of $CO_2$-concentration at a cooling-rate of 1.5 K/min and a second-order polynomial fit.**

However, his values of pressure sensitivity seems to be somewhat high, that might be due to higher cooling rate and a non-linearity between cooling rate and crystallization temperature at different saturation pressures. The fit parameters in dependence of $CO_2$-concentration, $c$, of the 2nd order polynomial $T_C = T_0 + B_1 c + B_2 c^2$ are: $T_0 = 428.31 \pm 1.5$~K, $B_1 = -1.366 \pm 0.229$ K/bar and $B_2 = 1.207 \cdot 10^{-2} \pm 6.52$~$10^{-3}$ K/bar$^2$.

**4. 4. Correlation of rheological data and crystallisation**

Since the crystallisation process is notably affected by the transport properties of the chains in the melt, the rheological properties were correlated to the crystallisation behavior. It can be seen that the shift factor is correlated to crystallisation temperature in an exponential relationship, as depicted in **Figure 11**.

An Arrhenius-analogue was used to quantify the correlation:

$$\ln(a_c(p,c)) = \frac{E_{a,c}}{R}\left(\frac{1}{T_{c,0}} - \frac{1}{T_c(p,c)}\right) \quad (10)$$

with the crystallization temperature at atmospheric conditions, $T_{C,0}$, the crystallization temperature at gas saturation under pressure $T_C(p,c)$ and an activation energy $E_{a,c}$. The activation energy is obtained by fitting, which yields $E_{a,c} = 142 \pm 7$ kJ/mol. This is three times the activation energy for flow from the Arrhenius equation obtained from oscillatory experiments, which is probably caused by the usage of the crystallization temperature instead of the glass transition temperature as by Royer et al. [33].

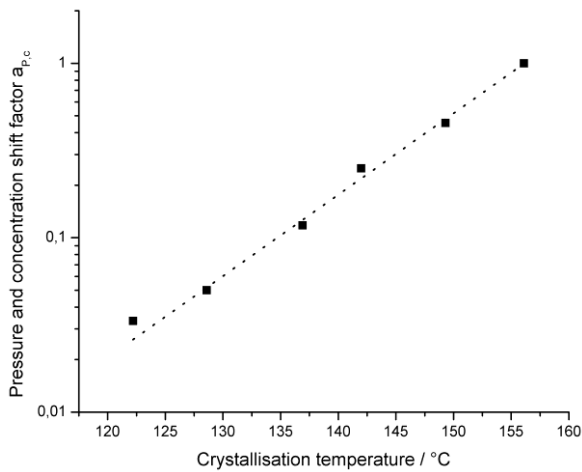

**Figure 11: Pressure and concentration shift factor dependence of the crystallisation temperature.**

These results emphasize the great potential of the pressure cell for measurements of crystallization phenomena and their correlation with rheological data. Besides the mentioned advantages for measuring the crystallization temperature of this rheological method over the currently used ones, it also enables one to obtain data of rheology and crystallization in one experiment.

Nevertheless, the specific errors of the new method must be considered. However, it will be reasoned that they are indeed negligible. For HPDSC and the pressure cell method the same principle error holds: temperature change with time leads to a changed solubility of the sample and thus to different $CO_2$-contents. However, this effect is localised to the outer edges of the sample and cannot reach the inner section due to long diffusion paths. Also, the gas diffusion rate is reduced during the measurement, since the polymer is cooled down, so that not much of the sample gets an increased $CO_2$-concentration. And since already a small crystallised part of the sample is enough to cause a big increase of torque this error is deemed negligible.

In contrast to HPDSC, this method is deemed superior, because the error of increasing $CO_2$-concentration during cooling is so small due to a relatively large sample size and thereby long diffusion times.

## 4.4 Comparison of the pressure cell and the in-line method

In order to validate the data of the in-line experiment under gas-loading, the results were compared to values from a previous publication [10] with a rotational rheometer MCR301 (Anton Paar, Austria) equipped with a coaxial plate-plate system in a pressure cell. In this publication, the detailed experimental procedure and further results from this method, with special emphasis on crystallisation phenomena, have been addressed by the authors [10].

At a CO$_2$-pressure of 50 bar the equation predicts a gas-concentration of 4.25 %, which is close to the concentration of 4 % used in the in-line experiment. **Figure 12** compares the results from the pressure cell and the in-line measurements. Interestingly, an appropriate agreement between both sets of data can be seen at two different shear rate scales.

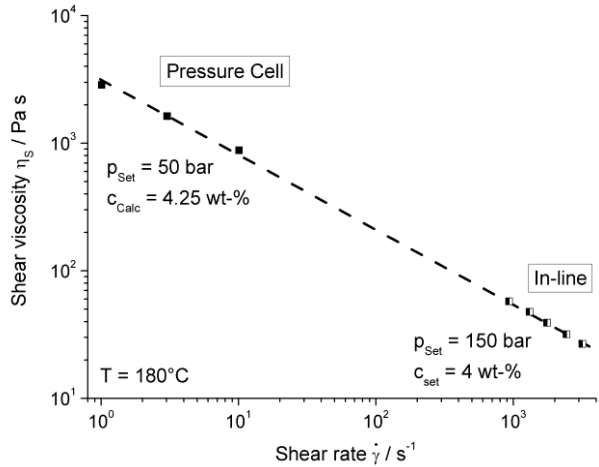

**Figure 12: Comparison of viscosity versus shear rate for gas-loaded LCB-PP as measured by the pressure cell and in-line rheometry at comparable CO$_2$-concentration. The broken line helps to guide the eye.**

## 5. Summary

An accurate estimation of the rheology in a gas-loaded polymer is crucial for precise controlling over polymer foaming processes. Rheology analysis is required for efficient die design and also can be used to achieve homogeneous cellular morphology of the foam, which in turn is a requirement for a multitude of applications such as thermal insulation. The dissolved gas can be mathematically contributed into formulating master-curves as well. Ideally, the contribution of the blowing agent is independent of the other variables of state (temperature and pressure). However, interdependencies exist in reality as was shown exemplarily in this chapter.

The most applied technique to study the effect of gas-loading on the rheological properties of a polymer melt is in-line rheometry. However, this method is rather expensive and requires large amounts of material to be tested. On the other side, the results can be interpreted rather accurate as the method has been established for a long time. Furthermore, realistic conditions in terms of shear rates can be accessed. Another method is using the pressure cell, which is an optional equipment for standard rotational rheometers. It allows the determination of diffusion, rheological and crystallization properties in a single experiment. However, this method is only applicable to high-viscosity polymers and the data evaluation can also be rather challenging. As a conclusion, every method has its pros and cons. The method of choice therefore has to be chosen according to the material properties, the already installed equipment, and the targeted application (i.e. the desired range of accessible shear rates). Eventually, exemplary results were shown for a long-chain branched (high melt strength) PP with both of the above-described methods. It is found that both approaches deliver comparable results and allow for a good evaluation of the material if used in an appropriate shear rate.

## Acknowledgments

The financial support of the German Research Foundation (DFG) in the frame of the research project number Al 474/18-1 is highly acknowledged. We thank Dr. Martin Laun and Dr. Sandra Aline Sanchez Vazquez for the fruitful discussions. We also would like to acknowledge the SFB 840. PP was kindly provided by Borealis AG, Austria.

19                                                                         A Chapter in *Polymer Rheology*